\documentclass[a4paper,fleqn,usenatbib]{mnras}

\usepackage{newtxtext,newtxmath}

\usepackage[T1]{fontenc}
\usepackage{ae,aecompl}

\usepackage{graphicx}	
\usepackage{amsmath}	
\usepackage{amssymb}	
\usepackage{booktabs}
\usepackage{listings}
\usepackage{xcolor}
\usepackage{breakurl}
\usepackage{microtype}

\defcitealias{FKP}{FKP}

\title[Detection of the BAO in the DR12 Bispectrum]{A Detection of the Baryon
Acoustic Oscillation Features in the SDSS BOSS DR12 Galaxy Bispectrum}

\author[D. W. Pearson \& L. Samushia]{
David W. Pearson,$^{1}$\thanks{E-mail: dpearson@phys.ksu.edu} \&
Lado Samushia$^{1,2,3}$
\\
$^{1}$Department of Physics, Kansas State University, 116 Cardwell Hall,
Manhattan, KS, 66506, USA\\
$^{2}$National Abastumani Astrophysical Observatory, Ilia State University, 2A
Kazbegi Ave., GE-1060 Tbilisi, Georgia\\
$^{3}$Institute of Cosmology \& Gravitation, University of Portsmouth, PO1 3FX,
UK }

\date{Accepted XXX. Received YYY; in original form 2017 December 12}

\pubyear{2017}

\begin{document}
\label{firstpage}
\pagerange{\pageref{firstpage}--\pageref{lastpage}}
\maketitle

\begin{abstract}
We present the first high significance detection ($4.1\sigma$) of the Baryon Acoustic Oscillations (BAO)
feature in the galaxy bispectrum of the twelfth data release (DR12) of the
Baryon Oscillation Spectroscopic Survey (BOSS) CMASS sample ($0.43 \leq z \leq
0.7$). We measured the scale dilation parameter, $\alpha$, using
the power spectrum, bispectrum, and both simultaneously for DR12, plus 2048
MultiDark-\textsc{patchy} mocks in the North and South Galactic
Caps (NGC and SGC, respectively), and the volume weighted averages of
those two samples (N+SGC). The fitting to the mocks validated our analysis
pipeline, yielding values consistent with the mock cosmology. By
fitting to the power spectrum and bispectrum separately, we tested the
robustness of our results, finding consistent values from the NGC,
SGC and N+SGC in all cases. We found $D_{\mathrm{V}} = 2032 \pm 24
(\mathrm{stat.}) \pm 15 (\mathrm{sys.})$~Mpc,
$D_{\mathrm{V}} = 2038 \pm 55 (\mathrm{stat.}) \pm 15 (\mathrm{sys.})$~Mpc, 
and $D_{\mathrm{V}} = 2031 \pm 22 (\mathrm{stat.}) \pm 10
(\mathrm{sys.})$~Mpc from the N+SGC power spectrum, bispectrum and simultaneous fitting, respectively. 
Our bispectrum measurement precision was mainly limited by the size of the covariance matrix.
Based on the fits to the mocks, we showed that if a less noisy estimator of the
covariance were available, from either a theoretical computation or a
larger suite of mocks, the constraints from the
bispectrum and simultaneous fits would improve to 1.1 per cent (1.3 per cent with systematics) and 
0.7 per cent 
(0.9 per cent with systematics), respectively, with the latter being slightly more precise than the power
spectrum only constraints from the reconstructed field.

\end{abstract}




\section{Introduction}

The bispectrum is sensitive to non-Gaussianities in the galaxy density field
from primordial physics, gravitational dynamics, velocity distortions and
biasing. However, bispectrum estimates are quite noisy since one can only
average over triangles of the same shape, but with different orientations. This
means that the number of coherence cells that contribute to a single bispectrum
measurement is relatively small \citep{Martinez2002}.  As the volume of our
surveys increases, the noise in the bispectrum should decrease, making it a
potentially powerful tool in improving constraints on cosmological parameters
\citep[see][equations (18) \& (19)]{Sefusatti2005}. In particular, future surveys such as
the Wide Field InfraRed Survey Telescope \citep[\textit{WFIRST};][]{wfirst} surveys
and the Dark Energy Spectroscopic Instrument (DESI) Bright Galaxy Survey \citep[BGS;][]{DESI}, will
simultaneously cover a large volume and have a high number density, making
the constraining power of the bispectrum comparable to that of the power spectrum
\citep{Gagrani2017}.

Recent studies, making use of the Sloan Digital Sky Survey's (SDSS) Baryon
Oscillation Spectroscopic Survey \citep[BOSS;][]{Dawson2013} data, have used
the galaxy bispectrum to help bolster constraints of various cosmological
parameters. \citet{Gil-Marin2015a} used the galaxy power spectrum and
bispectrum monopoles to constrain a model with two bias parameters, $b_{1}$ and
$b_{2}$, for a nonlinear, non-local bias model, along with the linear growth
parameter, $f$. It was the additional constraining power of the bispectrum that
allowed them to break the degeneracy between the bias and growth. More recently,
\citet{Gil-Marin2017} used the galaxy power spectrum monopole and quadrupole
along with the bispectrum monopole to perform a measurement of redshift-space
distortions (RSD).

The Baryon Acoustic Oscillation (BAO) peak postion allows us to constrain the 
expansion rate of our Universe by measuring
its size at a number of different redshifts. The measurement, in its simplest
form, involves constraining a single scale dilation parameter, $\alpha$, which
can then be related to the distance to a particular redshift 
\begin{equation}
\label{eq:alpha} 
\alpha \equiv
\dfrac{D_{\mathrm{V}}(z)r_{\mathrm{d}}^{\mathrm{fid}}}{D_{\mathrm{V}}^{\mathrm{fid}}(z)r_{\mathrm{d}}},
\end{equation} 
where $r_{\mathrm{d}}$ is the sound horizon at the drag epoch, and the
superscript `fid' refers to the fiducial cosmology. That is, the distance to
redshift $z$, is 
\begin{equation} 
\label{eq:DVmeas} 
D_{\mathrm{V}}(z) = \alpha
D_{\mathrm{V}}^{\mathrm{fid}}(z)\left(\dfrac{r_{\mathrm{d}}}
{r_{\mathrm{d}}^{\mathrm{fid}}}\right).
\end{equation}
This distance is also related to the angular diameter distance, and the Hubble
parameter at the redshift of interest via
\begin{equation}
\label{eq:DV}
D_{\mathrm{V}}(z) \equiv \left[cz(1+z)^{2}D_{\mathrm{A}}^{2}(z)H^{-1}(z)\right]^{1/3}.
\end{equation}
We can compute $H(z)$ simply as $H(z) = H_{0}E(z)$, with
\citep{Peebles1980,Martinez2002}
\begin{equation}
E^{2}(z) = \Omega_{\mathrm{M}}(1+z)^{3} + \Omega_{\Lambda} - \dfrac{kc^{2}}{H_{0}^{2}R_{0}^{2}}(1+z)^{2},
\end{equation}
and $k = 0$ for a spatially flat cosmology. The angular diameter distance is
given, in general, by
\begin{equation}
D_{\mathrm{A}}(z) = \dfrac{R_{0}}{1 + z}S_{k}(\omega(z)),
\end{equation}
where $R_{0}$ is the radius of curvature of the Universe, and 
\begin{equation}
S_{k}(\omega(z)) = \left\lbrace\begin{array}{ll}
\sin\omega(z), & k = 1, \\
\omega(z), & k = 0, \\
\sinh\omega(z), & k = -1.
\end{array}\right.
\end{equation}
So, in a flat universe, $S_{k}(\omega(z))$ is simply $\omega(z)$, which is
given by 
\citep[see][chapter 2]{Martinez2002}
\begin{equation}
\omega(z) = \dfrac{c}{R_{0}H_{0}}\int_{0}^{z}\dfrac{\mathrm{d}z'}{E(z')},
\end{equation}
giving the angular diameter distance as
\begin{equation}
D_{\mathrm{A}}(z) = \dfrac{c}{(1+z)H_{0}}\int_{0}^{z}\dfrac{dz'}{E(z')}.
\end{equation}
From this, given a fiducial cosmology and a constraint on $\alpha$, we can
easily calculate the distance, $D_{\mathrm{V}}(z)$.

When it comes to the BAO peak position, studies
tend to focus on two point statistics -- the correlation function and its
Fourier transform, the power spectrum \citep[see
e.g.][]{Anderson2012,Anderson2014,Cuesta2016,Gil-Marin2016,Ross2017,Beutler2017}.
This is justified by the fact that the reconstruction of the galaxy field
\citep{Eisenstein2007} is believed to at least partially `move' the
information from the three point statistics -- e.g. the three point correlation
function or the bispectrum -- into the two point statistics. However, measuring the BAO peak position
from the bispectrum directly -- or the combination of the non-reconstructed power spectrum and
bispectrum -- is still useful as the effect of the reconstruction on the
information content of the higher order statistics is, at the moment, not
completely clear. 

In a series of recent papers
\citep{Slepian2015,Slepian2017a,Slepian2017b} it was shown that the BAO peak
could be measured in the galaxy three point correlation function. Specifically, 
\citet{Slepian2017b}, using the SDSS BOSS Data Release 12 (DR12)
constant mass (CMASS) sample, were able to obtain a distance to redshift
$z=0.57$ with an accuracy of 1.7 per cent. Additionally, they conclude that the
three point correlation function contains significant additional information on
the distance scale.

Inspired by the results of \citet{Slepian2017b}, we set out to investigate
whether the same signal could be detected in the galaxy bispectrum. In this
paper, we present constraints on the scale dialiation parameter, $\alpha$, from the 
SDSS BOSS DR12 CMASS power
spectrum, bispectrum, and a simultaneous constraint from their combination. We
did this separately for the north galactic cap (NGC), south galactic cap (SGC)
and volume weighted north plus south galactic cap (N+SGC) samples for both the
data and the mock galaxy catalogues used. The mock results were used as
validation for our analysis pipeline and assessing systematic errors. We compared
our results with the analyses of
\citet{Anderson2014}, \citet{Cuesta2016}, \citet{Gil-Marin2016},
\citet{Ross2017} and \citet{Slepian2017b}, finding them to be consistent. 

Our constraint on $\alpha$ from the
bispectrum only fitting is weaker compared to the results presented in
\citet{Slepian2017b}. This is chiefly due to the number of mocks used for the 
covariance matrix estimation, being only about three times the number of data
points in the bispectrum measurement, which inflated our uncertainty
in the parameter. We show that if this additional uncertainty due to the
noisy covariance was negligible, our constraints on the distance scale would be
comparable to the ones derived in \citet{Slepian2017b}. When combined with the
power spectrum measurements the resulting constraint on $\alpha$ reached
one per cent and is therefore comparable with the constraints from the power
spectrum only of the reconstructed galaxy field presented in
\citet{Cuesta2016}. If not for the covariance matrix noise, the joint
power spectrum and bispectrum constraints would be slightly superior.

The layout of the paper is as follows: In Section \ref{sec:DataMeas} we briefly
discuss the data and mocks that were used in our analysis, and the procedures
used to measure the power spectrum, bispectrum and covariance. In Section
\ref{sec:ModFit} we present the models that were fitted to the data and outline
our fitting procedure. In Section \ref{sec:results} we present the main results
of this paper, and then provide some discussion in Section
\ref{sec:conclusion}.

Our fiducial cosmology was chosen to match the \textit{Planck} cosmic microwave
background results \citep{Planck2016}, with $\Omega_{\mathrm{M},0} =
0.3089$, $\Omega_{\Lambda,0} = 0.6911$, $\Omega_{\mathrm{b}}h^{2} = 0.02230$,
$h = 0.6774$, $n_{\mathrm{s}} = 0.9667$, and $\sigma_{8} = 0.8159$ \citep[i.e.
all values were taken from the last column of Table 4 in][]{Planck2016}.

\section{Data, Mocks \& Measurements}
\label{sec:DataMeas}

We used the BOSS DR12 CMASS sample containing 777 202 galaxies in the redshift
range $0.43 \leq z \leq 0.7$ \citep[see][for more details about this sample]{Alam2017}. 
This sample contains luminous red galaxies (LRGs)
selected as to ensure a roughly constant mass of all tracers. For the purposes
of covariance estimation and analysis pipeline validation, we made use of the
2048 MultiDark-\textsc{patchy} mock catalogues \citep{Kitaura2016,Rodriguez-Torres2016} released with DR12.

\subsection{Measuring the Power Spectrum}

To measure the galaxy power spectrum monopole, we used the standard
\citet*[][hereafter FKP]{FKP} method. To start, we calculated the weight for each galaxy using
the provided \citetalias{FKP} and systematic weights via the scheme suggested
by \citet{Anderson2012}, 
\begin{equation} 
\label{eq:weights} 
w(\bmath{r}) =
w_{\mathrm{FKP}}(\bmath{r})w_{\mathrm{sys}}(\bmath{r}) (w_{\mathrm{rf}} +
w_{\mathrm{cp}} - 1), 
\end{equation} 
where $w_{\mathrm{FKP}}$ are the usual \citetalias{FKP} weights,
$w_{\mathrm{sys}}$ are the combined systematics, $w_{\mathrm{rf}}$ and
$w_{\mathrm{cp}}$ are the redshift failure and close pair weights,
respectively, which are unity by default.  Next we binned the galaxies and
randoms onto a $512\times 1024\times 512$ grid, in a box of size $(1792\times
3584\times 1792)~h^{-3}~\mathrm{Mpc}^3$ using a cloud-in-cell (CIC)
interpolation scheme and the weights to give $n_{\mathrm{gal}}(\bmath{r})$ and
$n_{\mathrm{ran}}(\bmath{r})$, respectively.  Then we calculated the
over-density field 
\begin{equation} 
\delta(\bmath{r}) =
n_{\mathrm{gal}}(\bmath{r}) - \alpha_{\mathrm{den}}
n_{\mathrm{ran}}(\bmath{r}), 
\end{equation} 
where $\alpha_{\mathrm{den}}$ is the ratio of the sum of the galaxy weights to
the sum of the random weights.  This was then Fourier transformed using the
Fastest Fourier Transform in the West (\textsc{fftw}) library\footnote{
\href{http://fftw.org/}{fftw.org}}, and then averaged by frequency in bins of
width $\Delta k = 0.008$, 
\begin{equation}
\label{eq:Pkest} 
\hat{P}(k) = \dfrac{1}{\int
\bar{n}^{2}(\bmath{r})w^{2}(\bmath{r})}
\left[\left\langle|\delta(\bmath{k})|^{2}\right\rangle -
S_{\mathrm{P}}\right]G^{2}(\bmath{k}).  
\end{equation} 
Here $\bar{n}(\bmath{r})$ is the average number density at $\bmath{r}$,
$S_{\mathrm{P}}$ is the shot-noise defined by 
\begin{equation} 
S_{\mathrm{P}} =
\int_{\mathrm{gal}} \bar{n}(\bmath{r})w^{2}(\bmath{r})\mathrm{d}\bmath{r} +
\alpha_{\mathrm{den}} \int_{\mathrm{ran}}
\bar{n}(\bmath{r})w^{2}(\bmath{r})\mathrm{d}\bmath{r}, 
\end{equation} 
and $G(k)$ is the CIC binning correction \citep{Jeong2010}, 
\begin{equation}
G(\bmath{k}) =
\left[\mathrm{sinc}\left(\dfrac{k_{x}L_{x}}{N_{x}}\right)\mathrm{sinc}\left(\dfrac{k_{y}
L_{y}}{N_{y}}\right)\mathrm{sinc}\left(\dfrac{k_{z}L_{z}}{N_{z}}\right)\right]^{-2},
\end{equation} 
where $\mathrm{sinc}(x) = \sin(x)/x$, $L_{i}$ is the length of the box you
placed your sample into in the $i$ direction, and $N_{i}$ is the number of grid
cells in that dimension. We did not use a reconstruction procedure
\citep{Eisenstein2007} on our over-density field, as this further correlates
the power spectrum and bispectrum measurements \citep[see][section
8.2]{Slepian2017b} and removes most of the bispectrum signal. When combining
the NGC and SGC, because of the large size of the combined power spectrum and bispectrum
covariance matrix, we used a
simple volume weighted averaging instead of a potentially more efficient
inverse covariance matrix based weighting.

\begin{figure} 
\includegraphics[width=\linewidth]{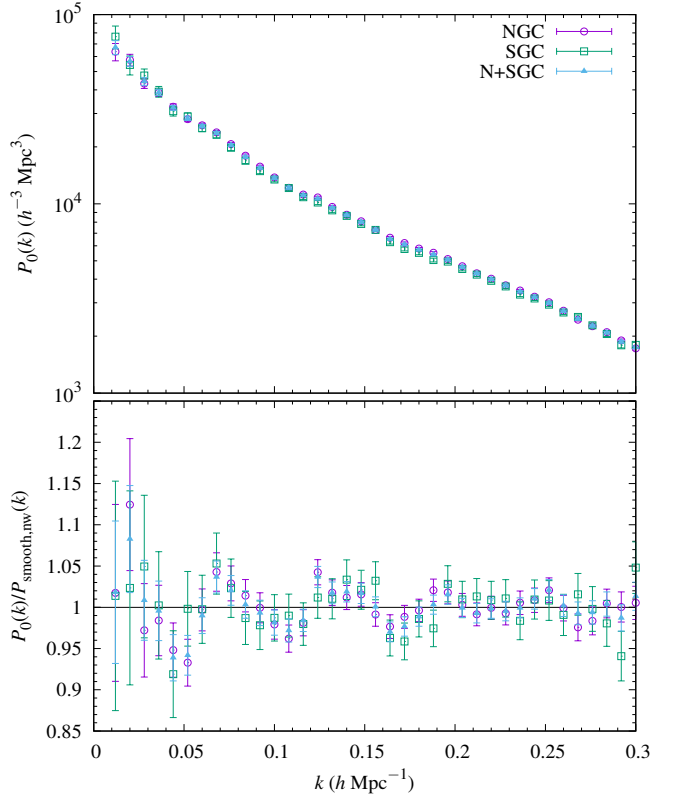}
\caption{The measured power spectrum from the NGC, SGC, and their volume
weighted average (top panel). To better see how they compare, they are also
shown normalized by $P_{\mathrm{smooth,nw}}(k)$ as defined in equation
\eqref{eq:smooth} in the bottom panel (See the online article for a colour
version of this plot.)} 
\label{fig:Pkdata} 
\end{figure}

First, we calculated the effective survey volumes of the NGC and SGC via
\citep{Anderson2012}, 
\begin{equation} 
V_{\mathrm{eff}} = \sum_{i}
\left(\dfrac{\bar{n}(z_{i})P_{0}}{1 + \bar{n}(z_{i})P_{0}}\right)^{2} \Delta
V(z_{i}), 
\end{equation} 
where $P_{0}$ is the approximate power spectrum amplitude that the
\citetalias{FKP} weights were optimizing for, $\bar{n}(z_{i})$ is the average
number density of galaxies in the redshift bin $z_{i}$, and $\Delta V(z_{i})$
is the volume of the spherical shell at that redshift scaled by the fraction of
the sky covered by the survey. The weights were then 
\begin{equation} 
\label{eq:volweights} 
w_{\mathrm{V},i} =
\dfrac{V_{\mathrm{eff},i}}{V_{\mathrm{eff,NGC}} + V_{\mathrm{eff,SGC}}},
\end{equation} 
where $i$ was either NGC or SGC. The combined power spectrum measurement was
then 
\begin{equation} 
\label{eq:Pkcomb}
\hat{P}_{\mathrm{N+SGC}}(k) = w_{\mathrm{V,NGC}}\hat{P}_{\mathrm{NGC}}(k) +
w_{\mathrm{V,SGC}}\hat{P}_{\mathrm{SGC}}(k).  
\end{equation} 
The measured power spectra are shown in Figure \ref{fig:Pkdata}, along with the
measurements normalized by equation \eqref{eq:smooth} calculated with the best
fitting parameter values. All of the measurements agree with each other quite
well. 

\subsection{Measuring the Galaxy Bispectrum} 

We used the length of the three wave-vectors, $k_1$, $k_2$, and $k_3$, restricted by the
triangle condition, $\bmath{k}_{1} + \bmath{k}_{2} + \bmath{k}_{3} = 0$, to
parametrize the shape of the bispectrum and averaged over two angles describing
the orientation of the triangle \citep[see e.g.][]{Gil-Marin2015a,Gagrani2017}.
Since we were only using the monopole of the bispectrum we were able to use 
a Cartesian fast Fourier transform (FFT) without having to worry about the 
wide-angle effects \citep{Samushia2015,Scoccimarro2015}.

We employed a `brute-force' algorithm that explicitly looked at all triplets that
fell within a specific ($k_1, k_2, k_3$) bin and satisfied the triangle condition.
This algorithm scales as $\mathcal{O}(N_{k}^{2})$ -- where $N_{k}$ is the
number of grid points in Fourier space where $k$ is in the range of interest. 

It is possible to calculate the bispectrum using point-wise products of FFTs 
of shells of $\delta(k_{1})$, $\delta(k_{2})$ and
$\delta(k_{3})$, with $\delta(k)$ being the FFT of the galaxy over-density
field \citep[see e.g.][section 5.1]{Baldauf2015}. However, we found that it was
computationally more efficient to use the `brute-force' algorithm implemented
on a graphics processor unit (GPU)\footnote{We suspect that a GPU based
implementation of the shell FFT method would be even faster than our
$\mathcal{O}(N_{k}^{2})$ method, but could not verify this given the memory
limitations (2 GB) of the GPU used for this work.}.

Our estimate of the bispectrum was \citep{Scoccimarro2000,Scoccimarro2001}
\begin{equation}
\hat{B}(k_1, k_2, k_3) = \dfrac{\left\lbrace\langle\Re[
\delta(\bmath{k}_{1})\delta(\bmath{k}_{2})\delta(\bmath{k}_{3})]\rangle - S_{\mathrm{B}}\right\rbrace
G(\bmath{k}_{1})G(\bmath{k}_{2})G(\bmath{k}_{3})}{\int \bar{n}^3(\bmath{r})w^3(\bmath{r})\mathrm{d}\bmath{r}}
\end{equation}
where $\Re$ denotes the real part of a complex number and $\delta(\bmath{k})$
is the Fourier transformed over-density field used in the power spectrum
calculation. The shotnoise $S_{\mathrm{B}}$, was computed using the power
spectrum estimate of equation \eqref{eq:Pkest}
\citep{Scoccimarro2000,Scoccimarro2001}
\begin{equation}
S_{\mathrm{B}} = (\hat{P}_1 + \hat{P}_2 + \hat{P}_3)\int \bar{n}^{2}
(\bmath{r})w^{3}(\bmath{r})\mathrm{d}
\bmath{r} + (1 - \alpha^{2})
\int \bar{n}(\bmath{r})w^{3}(\bmath{r})\mathrm{d}
\bmath{r},
\end{equation}
where $\hat{P}_{i} = \hat{P}(k_{i})$ is the estimate of the power spectrum at
$k_{i}$. We computed the combined N+SGC bispectrum in the same way as we
computed combined power spectrum, using the weights of equation
\eqref{eq:volweights} and simply replacing the power spectrum measurements in
equation \eqref{eq:Pkcomb} with the bispectrum measurements.

To implement our estimator on the GPU, we first trimmed the Fourier transformed
over-density field keeping only the wave numbers in our range of interest, $0.04
\leq k < 0.168$, to better fit in the limited memory resources of the GPUs used
for the calculation. We then generated an array of $\bmath{k}$ values stored as
integer multiples of the fundamental frequency in each coordinate direction.
Each GPU thread took a single $\bmath{k}_{1}$ and looped over the other
$\bmath{k}$'s as $\bmath{k}_{2}$, starting at $\bmath{k}_{1}$ and going through
vectors not yet used as $\bmath{k}_{1}$, to avoid double counting. The value of
$\bmath{k}_{3}$ was then computed from the triangle condition and checked to
ensure it was in the range of interest. 

For convenience and speed, we used the
\textsc{cuda} types \texttt{int4} and \texttt{float4} (mocks) or
\texttt{double4} (data) for storing the $\bmath{k}$'s and the over-density
field, respectively, allowing for four numbers of each type to be stored at a
single array index. The \texttt{int4} type allowed us to store the three
components as well as the corresponding grid index for the over-density field,
which made the lookup times negligible.  For the over-density field, the first
two values were the real and imaginary parts, the third value was the magnitude
of $\bmath{k}$ at that grid point, and the fourth value was the correction for
the CIC binning associated with that point.

As the GPUs used for this work had relatively low double-precision performance
-- really not much better than a typical CPU -- to achieve high-throughput
while retaining accuracy it was necessary to use mixed precision for
calculating the bispectrum from the 2048 NGC and 2048 SGC mocks.  Effectively,
the calculation of $\bmath{k}_{3}$ was done as integer math, the overall
grid-correction was calculated using single-precision, while the bispectrum
contribution was calculated using single-precision, but stored and then binned
as double-precision. The binning was done via a \textsc{cuda} function
\texttt{atomicAdd}. While the double-precision version of this function is only
supported on NVidia GPUs of compute capability 6 or higher, the documentation
provides an implementation that can be used on GPUs of lower compute capabilities. 

The reason it is not implemented on those older GPUs, however, is due to its
relatively low performance, meaning that it was by far the most expensive step
in the process. To reduce the impact of this step, the binning was first done
in the GPU thread block shared memory, which is much lower latency than the global
GPU memory. Once all of the threads in a block went through all of their $k_{2}$'s, the histogram
was then binned into global memory in parallel. Naively, it may seem that the two step
process would be less efficient, however, due to the lower latency of the
shared memory, reduction in number of writes to global memory, and fewer bank
conflicts, you can often see at least a factor of 2 speed up
\citep{Sakharnykh2015}.

Implemented in this manner, we were able to compute the bispectrum from a
single mock in $\sim$94~s, which was 15 to 20 times faster than our FFT based
implementation, depending on whether a newer Intel core i7 or older AMD FX
processor was used, respectively. Our full double precision implementation\footnote{With 
the recent release of the 
NVidia Titan V, this time could be reduced to be comparable with the typical time
to calculate the power spectrum, e.g. $\sim$3~s, for between 1/3 and 1/6 the cost
of the comparable, but slightly faster NVidia Tesla V100. This will make studies of the 
bispectrum substantially quicker in the future.} used
for processing the data took $\sim$310~s. If we had used that implementation to
process the mocks, it would have taken approximately 14.5 days to complete. Our
mixed precision implementation cut that down to about 4.5 days.

\begin{figure*}
\includegraphics[width=\linewidth]{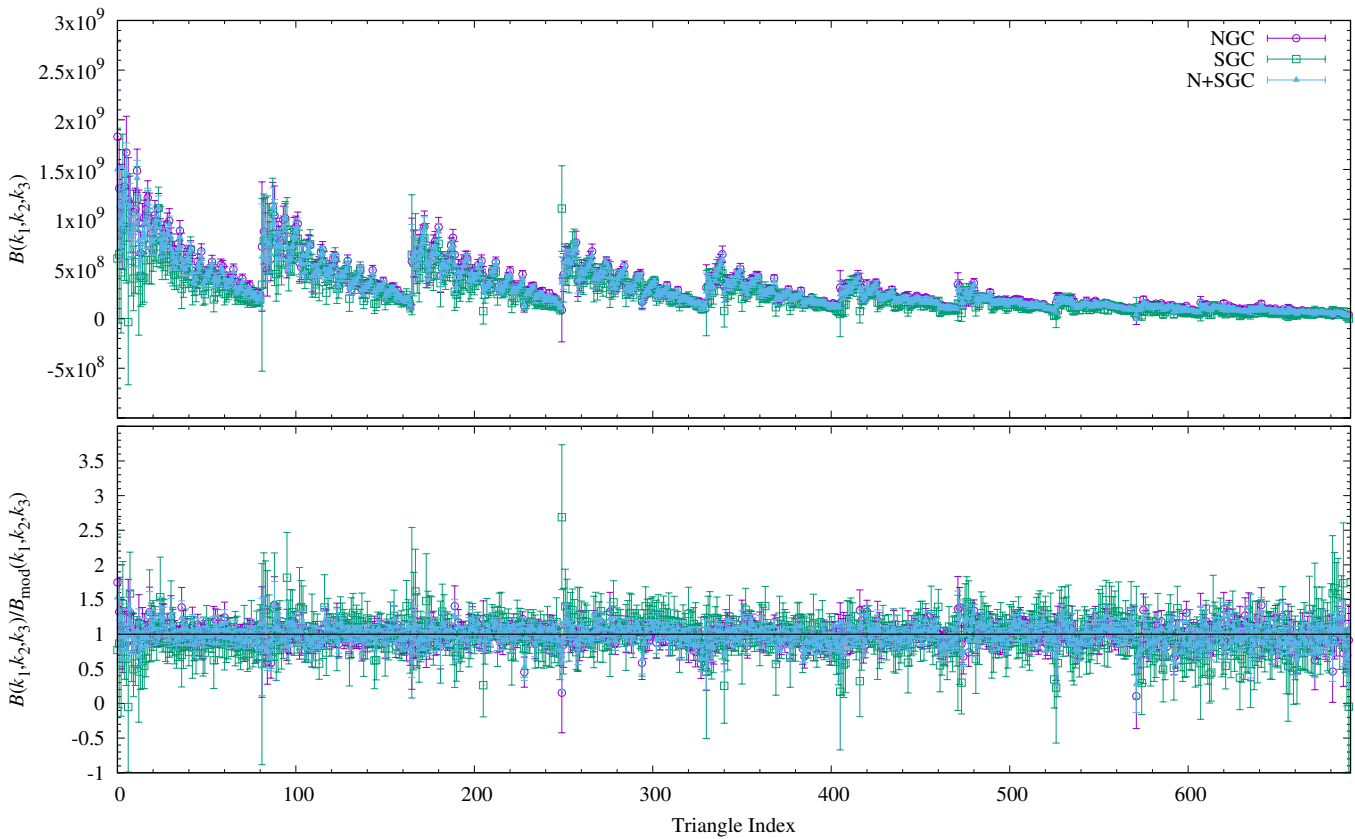}
\caption{The measured bispectrum from the NGC, SGC, and N+SGC samples. The top
panel shows the measurements with uncertainties ($\sqrt{C_{ii}}$) using the same
symbols (and colours for the online version) as in Figure \ref{fig:Pkdata}. The
bottom panel shows the data normalized by the best fitting model. The fact that
the normalized data are simply scattered about one shows that our model
accounts for the non-linearities in the data quite well.}
\label{fig:Bispec}
\end{figure*}

We show the measured bispectrum from the NGC, SGC and N+SGC samples in Figure
\ref{fig:Bispec}. We also show the measurements normalized by our best fitting
model -- see section \ref{sec:BispecMod} -- to show that it is able to
fit the data well. 

\begin{figure}
\includegraphics[width=\linewidth]{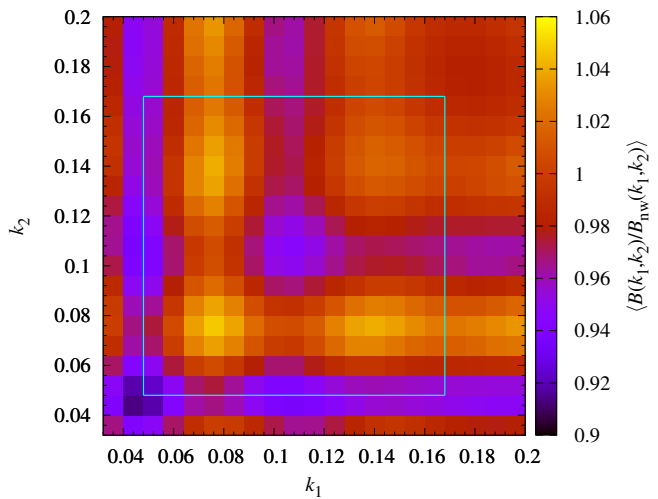}
\caption{The theoretical bispectrum normalized by the theoretical no-wiggle bispectrum,
as a function of the length
of two wave-vectors, averaged over the length of the third wave-vector. The inset box (cyan
in the online version)
encloses the scales used in our analysis. The two dimensional sequence of local
maxima and minima are manifestations of the BAO feature. (See the online  article 
for a colour version of this plot.)}
\label{fig:Bispec2D}
\end{figure}

Unlike the power spectrum, the raw bispectrum measurements do not visibly show the BAO feature. 
To elucidate the type of signal our constraints come from, we provide a two-dimensional plot of the
theoretical bispectrum calculated from a BAO power spectrum normalized by one calculated from a no-wiggle 
power spectrum as a function of two wave-vector magnitudes, averaged over the length of the third, in Figure 
\ref{fig:Bispec2D} (see section \ref{sec:BispecMod}). Plotted in this manner, it is possible to see clear 
hills and valleys coming from the BAO feature which are equivalent to the decaying oscillatory signature in 
the power spectrum. 

Additionally, we used this plot, along with the need to keep the data vector small 
enough to ensure reasonable covariance matrices, to decide what bispectrum wave number range to use in our 
fits. The inset square (cyan in the online version) shows our selected region, which should include the
equivalent of the first two 
`wiggles' in the power spectrum, helping to maximize the BAO constraining power, while keeping the data 
vector reasonably sized. We note that this two-dimensional plot is merely a convenient way of displaying the 
BAO features in the bispectrum. Our actual constraints on the scale dilation parameter came from fitting to 
the three dimensional data shown in Figure \ref{fig:Bispec}.

\subsection{Covariance}

We computed the sample covariance from the 2048 MultiDark-\textsc{patchy} mocks provided with DR12
\citep{Kitaura2016,Rodriguez-Torres2016}.  This was the main limit to the number of triangles we
could use for fitting the bispectrum data. First, we needed to estimate the covariance matrix
to enough accuracy that it was not singular, if we were to invert it for our
maximum likelihood fitting. Additionally, the errors in our covariance matrix
carry through and affect our constraints on the model parameters. To ensure
that the matrix was invertible, and that the uncertainties of its elements were
kept low, we limited ourselves to $0.04 \leq k_{1},k_{2},k_{3} \leq 0.168$ for
the bispectrum measurements, and $0.008 \leq k < 0.304$ for the power spectrum.
In total, we had 691 bispectrum triangles and 37 power spectrum points, which
gave at most 728 elements in our data vector.

\begin{figure*}
\includegraphics[width=\linewidth]{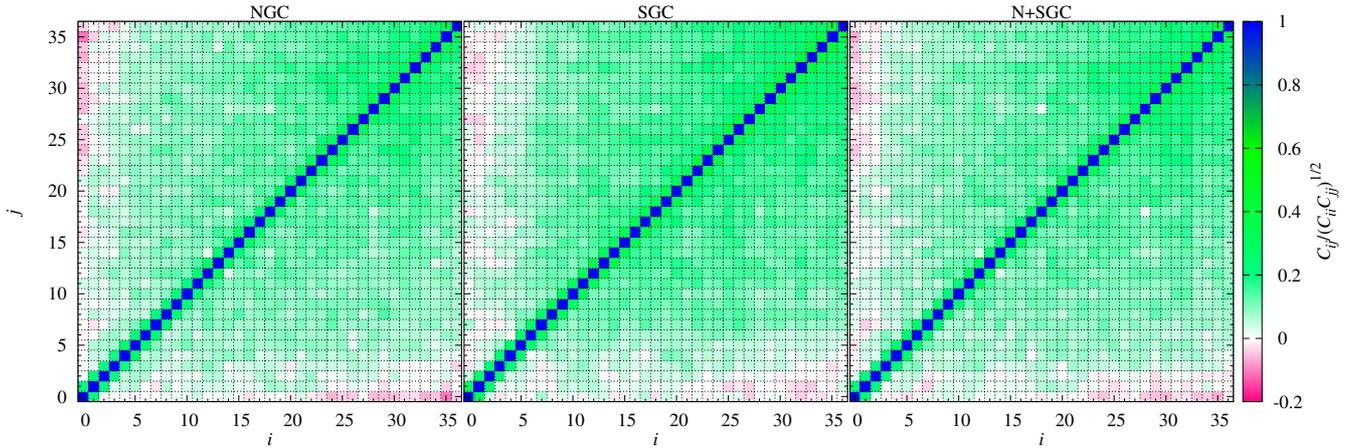}
\caption{The correlation matrices for the power spectrum measurements calculated
from 2048 Patchy mocks. The power spectra were calculated from the NGC and SGC,
then a weighted average. Most of the off-diagonal values vary between $\pm
0.2$. (See the online article for a colour version of this plot.)}
\label{fig:Pk_cor}
\end{figure*}

\begin{figure*}
\includegraphics[width=\linewidth]{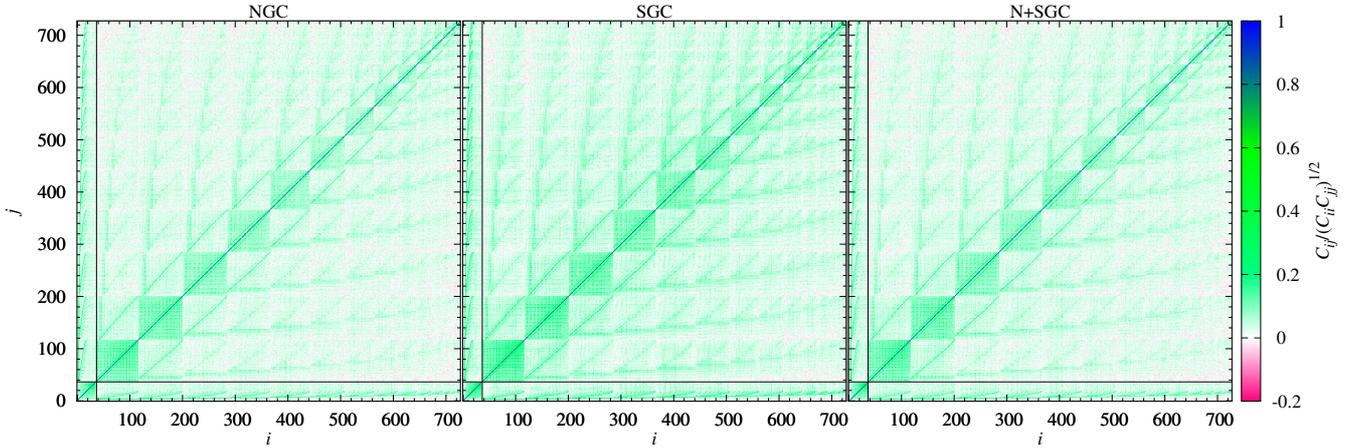}
\caption{The correlation matrices for the power spectrum and bispectrum with
cross-correlations calculated from the 2048 Patchy mocks. The vertical and
horizontal solid lines divide the matrix into the power spectrum (bottom left,
small square), the bispectrum (top right, large square), and their
cross-correlations (vertical rectangle on the left, or horizontal rectangle on
the bottom). The off-diagonal elements still mostly only vary between $\pm
0.2$. (See the online article for colour a version of this plot.)}
\label{fig:PkBk_cor}
\end{figure*}

We ran all of the mocks through our data pipelines as described above, then
computed the sample covariance,
\begin{equation}
C_{ij} = \dfrac{1}{n_{s} - 1}\sum_{s} (x_{i}^{s} - \bar{x}_{i})(x_{j}^{s} - \bar{x}_{j}),
\end{equation}
where $n_{s}$ is the number of mock samples, and the $s$ in the sum refers to a specific
sample. We did this for the NGC, SGC and N+SGC samples with the power spectrum, bispectrum
and both combined as our data vector, $\bmath{x}$. We plot the resulting
correlation matrices, $r_{ij} = C_{ij}/(C_{ii}C_{jj})^{1/2}$, in Figures
\ref{fig:Pk_cor}, and \ref{fig:PkBk_cor}. All seem reasonably well behaved,
with more than 90 per cent of the off-diagonal values falling between $\pm
0.2$.

When using the covariance matrix for our parameter fitting, we corrected for
the fact that the inverse of the covariance matrix is a biased estimate of the
true inverse covariance needed. This correction was simply
\citep{Hartlap2007,Percival2014}
\begin{equation}
\Psi = \left(\dfrac{n_{s} - n_{b} - 2}{n_{s} - 1}\right)\bmath{\mathsf{C}}^{-1}
\end{equation}
where $n_b$ is the number of values
in our data vector. Given that $n_{s} = 2048$, and $n_{b} = 37,~691$, and $728$
for the power spectrum, bispectrum, and combined data vectors, respectively,
the above correction factor was then 0.981, 0.662, and 0.644, meaning our
constraints on parameters from the bispectrum and the combined data were
affected by the limited number of mock catalogues.

In addition to the correction above, we also had to carry through the
uncertainty in the covariance matrix elements themselves due to limited number
of mock catalogues \citep{Taylor2013}. This affected the variance of the
parameters being constrained, increasing them by a factor of
\citep{Percival2014}
\begin{equation}
\label{eq:m1}
m_{1} = \dfrac{1 + \beta(n_{b} - n_{p})}{1 + A + \beta(n_{p} + 1)}, 
\end{equation}
where
\begin{equation}
A = \dfrac{2}{(n_{s} - n_{b} - 1)(n_{s} - n_{b} - 4)},
\end{equation}
\begin{equation}
\beta = \dfrac{(n_{s} - n_{b} - 2)}{(n_{s} - n_{b} - 1)(n_{s} - n_{b} - 4)},
\end{equation}
and $n_{p}$ is the number of parameters in the model being fit to the data. For
our analysis we found values of $m_{1} = 1.0089,~1.4982$, and $1.5248$ for the power
spectrum, bispectrum and combined sample, respectively. This substantially
reduced the potential constraints from the bispectrum and the combined samples,
highlighting a need for either a larger number of mock catalogues or a high
precision theoretical model of the covariance.

\section{Model \& Fitting}
\label{sec:ModFit}

\subsection{The Power Spectrum}
\label{sec:PkMod}

\begin{figure}
\centering
\includegraphics[width=\linewidth]{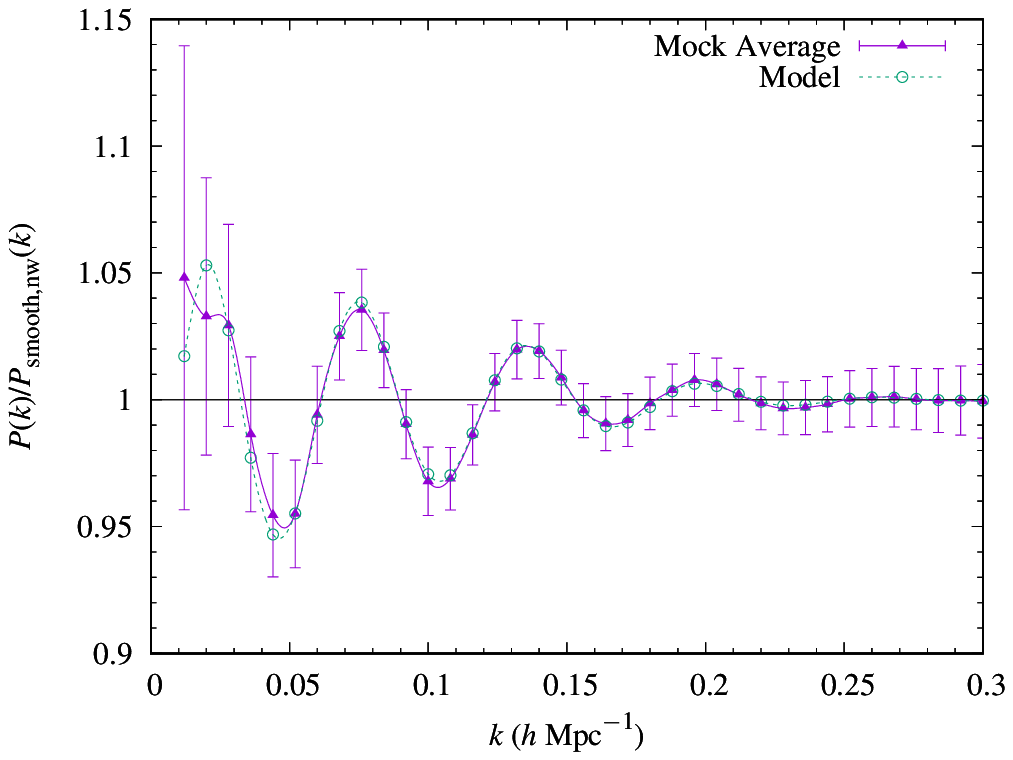}
\caption{A comparison of the best fitting model and the average measured power
spectrum from mock galaxy catalogues. The triangles show the mock average
normalized by $P_{\mathrm{smooth,nw}}(k)$ defined in equation
\eqref{eq:smooth}, calculated with the best fitting values of the $a_{i}$, and
$B$ parameters, with a smooth cubic spline drawn through the points. The open
circles show the model of equation \eqref{eq:PkMod} with the best fitting
parameters normalized again by $P_{\mathrm{smooth,nw}}(k)$. The two agree quite
well, particularly with respect to the BAO peak positions.}
\label{fig:PkModComp}
\end{figure}

To model the power spectrum, we followed the method of \citet{Anderson2014},
with one modification. We used the Python implementation of \textsc{camb}
\citep{Lewis2000} to generate a linear power spectrum with our fiducial
cosmology. Then, using the fitting formula of \citet{Eisenstein1998}, we
calculated a no-wiggle power spectrum to match the broadband shape of the
linear power spectrum from \textsc{camb}. Since the data contained
non-linearities, we used a smoothing polynomial to match the linear theory
power spectrum to the data,
\begin{equation}
\label{eq:smoothingPoly}
\mathcal{P}(k) = a_{0}k^2 + a_{1}k + a_{2} + \dfrac{a_{3}}{k} + \dfrac{a_{4}}{k^2} + \dfrac{a_{5}}{k^{3}}.
\end{equation}
This differs from the method used in \citet{Anderson2014} by including a term
proportion to $k^{2}$. We found that including this extra term gives us a
better fit. We added this polynomial to the no-wiggle power spectrum
\begin{equation}
\label{eq:smooth}
P_{\mathrm{smooth,nw}}(k) = B^{2}P_{\mathrm{nw}}(k) + \mathcal{P}(k),
\end{equation}
where $B$ is an amplitude parameter to account for galaxy bias and gravitational
growth. This was then multiplied by an oscillatory part
\begin{equation}
\label{eq:oscillate}
O(k) = \left[1 + \left(\dfrac{P_{\mathrm{lin}}(k/\alpha)}{P_{\mathrm{nw}}(k/\alpha)} - 1\right)\exp\left(\dfrac{1}{2}\Sigma^{2}k^{2}\right)\right]
\end{equation}
where $\alpha$ is defined in equation \eqref{eq:alpha}, and $\Sigma$ is a
Finger-of-God damping parameter. This gave our final
power spectrum model as
\begin{equation}
\label{eq:PkMod}
P_{\mathrm{mod}}(k) = P_{\mathrm{smooth,nw}}(k)O(k).
\end{equation}
Figure \ref{fig:PkModComp} shows the average power spectrum of the 2048 mocks
divided by the no-wiggle best fitting model -- e.g. the average divided by
equation \eqref{eq:smooth} with the best fitting values of the parameters -- as
well as equation \eqref{eq:oscillate} with the best fitting parameter values.

For the fitting, this model had 9 free parameters, the $a_{i}$ of equation
\eqref{eq:smoothingPoly}, the amplitude parameter, $B$, the non-linear damping
parameter, $\Sigma$, and the scale dilation parameter, $\alpha$. We used a
custom Markov Chain Monte Carlo (MCMC) code which utilized the Metropolis-Hastings
algorithm \citep{Hastings1970}. This choice was made so that when fitting
simultaneously to the power spectrum and bispectrum, the exact same code and
algorithm was used for both, as the bispectrum fitting had to be done with custom
code in order to utilize a GPU to speed up the model calculation (see section \ref{sec:BispecMod}).

\subsection{The Bispectrum}
\label{sec:BispecMod}

To model the galaxy bispectrum we used the second-order perturbation theory
model presented by \citet{Scoccimarro2000}, with some small changes to account
for non-linearities in the data. The first and second-order kernels are
\begin{equation}
\label{eq:1stkernel}
Z_{1}(\bmath{k}) = (b_{1} + f\mu^2)
\end{equation}
and
\begin{equation}
\label{eq:2ndkernel}
\begin{array}{@{}l@{\:}l}
Z_{2}(\bmath{k}_{1},\bmath{k}_{2}) =& \dfrac{b_{2}}{2} + b_{1}F_{2}(\bmath{k}_{1},\bmath{k}_{2})
+ f\mu^{2}G_{2}(\bmath{k}_{1},\bmath{k}_{2}) \\[1ex]
&+ \dfrac{f\mu k}{2}\left[
\dfrac{\mu_{1}}{k_{1}}Z_{1}(\bmath{k}_{2}) + \dfrac{\mu_{2}}{k_{2}}Z_{1}(\bmath{k}_{1})\right],
\end{array}
\end{equation}
where $\mu \equiv \bmath{k}\cdot\hat{\bmath{z}}/k$, with $\bmath{k} \equiv \bmath{k}_{1} + \bmath{k}_{2}$,
$\mu_{i} \equiv \bmath{k}_{i}\cdot\hat{\bmath{z}}/k_{i}$,
\begin{equation}
\label{eq:Fkernel}
F_{2}(\bmath{k}_{1},\bmath{k}_{2}) = \dfrac{5}{7} + \dfrac{\bmath{k}_{1}\cdot\bmath{k}_{2}}{2k_{1}k_{2}}
\left(\dfrac{k_{1}}{k_{2}} + \dfrac{k_{2}}{k_{1}}\right) + \dfrac{2}{7}
\left(\dfrac{\bmath{k}_{1}\cdot\bmath{k}_{2}}{k_{1}k_{2}}\right)^{2},
\end{equation}
and
\begin{equation}
\label{eq:Gkernel}
G_{2}(\bmath{k}_{1},\bmath{k}_{2}) = \dfrac{3}{7} + \dfrac{\bmath{k}_{1}\cdot\bmath{k}_{2}}{2k_{1}k_{2}}
\left(\dfrac{k_{1}}{k_{2}} + \dfrac{k_{2}}{k_{1}}\right) + \dfrac{4}{7}
\left(\dfrac{\bmath{k}_{1}\cdot\bmath{k}_{2}}{k_{1}k_{2}}\right)^{2}.
\end{equation}
From these, our model bispectrum was
\begin{equation}
\label{eq:bkmod}
\begin{array}{@{}l@{\;}l}
B(\bmath{k}_{1},\bmath{k}_{2},\bmath{k}_{3}) =& [2Z_{2}(\bmath{k}_{1},\bmath{k}_{2})Z_{1}(\bmath{k}_{1})
Z_{1}(\bmath{k}_{2})P(k_{1})P(k_{2})\\
& + \;\mathrm{cyc.}] D_{\mathrm{FoG}}(\bmath{k}_{1},\bmath{k}_{2},\bmath{k}_{3}),
\end{array}
\end{equation}
where
\begin{equation}
\label{eq:BispecFoG}
D_{\mathrm{FoG}}(\bmath{k}_{1},\bmath{k}_{2},\bmath{k}_{3}) = \dfrac{1}{1 + (k_{1}^{2}\mu_{1}^{2}
+ k_{2}^{2}\mu_{2}^{2} + k_{3}^{2}\mu_{3}^{2})\sigma_{v}^{2}/2}
\end{equation}
and $P(\bmath{k})$ is the non-linear power spectrum calculated for our fiducial cosmology
from \textsc{camb}. The addition of the
Finger-of-God suppression factor and the use of a non-linear power spectrum
were the changes we made to better fit the data. Even though this model has
been shown to be inadequate for fitting the full nonlinear bispectrum shape
\citep{Gil-Marin2015a,Gil-Marin2017} we found it to be adequate for the range of $k$ 
that we consider (e.g. $0.04 \leq k \leq 0.168$) in our analysis.
Since we were only interested in the position of the BAO peak we could afford to
marginalize over smooth systematic effects in the bispectrum shape without
properly modelling them. We found
that this was enough to get unbiased estimates of $\alpha$ and no extra smoothing
polynomials seemed to be necessary to achieve a good fit to the bispectrum data.

Since we were examining the bispectrum monopole, equation \eqref{eq:bkmod} was
spherically averaged by integrating over two angles: the angle of
$\bmath{k}_{1}$ with the line of sight, $\mu_{1}$, and the azimuthal angle of
$\bmath{k}_{2}$ around $\bmath{k}_{1}$, $\phi$ \citep[see e.g.][section
3.1]{Gagrani2017}. This gave the model as
\begin{equation}
\label{eq:bkmodint}
B(k_1,k_2,k_3) =\dfrac{1}{4\pi}\int_{-1}^{1}\mathrm{d}\mu_{1}\int_{0}^{2\pi}\mathrm{d}\phi
B(\bmath{k}_{1},\bmath{k}_{2},\bmath{k}_{3}).
\end{equation}
The values of $\mu_{2}$ and $\mu_{3}$ could easily be calculated from the value
of the two angles above.  We took the Alcock-Paczynski effect
\citep{Alcock1979,Kaiser1987,Ballinger1996,Simpson2010,Samushia2011} into
account by transforming the measured $k$'s and $\mu$'s as
\begin{equation}
k_{i} \longrightarrow \dfrac{k_{i}}{\alpha_{\perp}}\left[1 + \mu_{i}^{2}\left(\dfrac{\alpha_{\perp}^{2}}
{\alpha_{\parallel}^{2}} - 1\right)\right]^{1/2},
\end{equation}
\begin{equation}
\mu_{i} \longrightarrow \dfrac{\alpha_{\perp}}{\alpha_{\parallel}}\mu_{i}\left[1 + \mu_{i}^{2}
\left(\dfrac{\alpha_{\perp}^{2}}{\alpha_{\parallel}^{2}} - 1\right)\right]^{-1/2},
\end{equation}
where
\begin{equation}
\label{eq:alpha_para}
\alpha_{\parallel} = \dfrac{H^{\mathrm{fid}}(z)r_{\mathrm{d}}^{\mathrm{fid}}}
                    {H(z)r_{\mathrm{d}}},
\end{equation}
and
\begin{equation}
\label{eq:alpha_perp}
\alpha_{\perp} = \dfrac{D_{A}(z)r_{\mathrm{d}}^{\mathrm{fid}}}
                       {D_{A}^{\mathrm{fid}}(z)r_{\mathrm{d}}},
\end{equation}
along with renormalizing the power spectrum by a factor of
$1/\alpha_{\perp}^{2}\alpha_{\parallel}$ and due to equation \eqref{eq:bkmod},
the bispectrum by the same factor squared. From equations \eqref{eq:alpha},
\eqref{eq:DV}, \eqref{eq:alpha_para} and \eqref{eq:alpha_perp}, it can be seen
that $\alpha_{\parallel}$ and
$\alpha_{\perp}$ are related to $\alpha$ via
\begin{equation}
\label{eq:alphacombo}
\alpha^{3} = \alpha_{\perp}^{2}\alpha_{\parallel}.
\end{equation}

Since the double-integral of equation \eqref{eq:bkmodint} had to be evaluated
for each of our 691 $k$-triplets a very large number of times for the MCMC
fitting procedure, it had to be implemented in a
numerically efficient manner. For this we again turned to the GPU allowing us
to calculate the double-integral using a couple of levels of parallelism.

While it is possible to implement adaptive quadrature on the GPU, the error
estimation steps can introduce significant overhead, and most algorithms rely
on recursion which is not well suited for GPUs \citep{Thuerck2014}. Given this,
we instead opted for the much easier to implement, fixed Gaussian quadrature
rules. Since Gaussian quadrature can give an exact result for polynomials of
degree $2n - 1$ or less, it can allow very accurate numerical integration with
relatively few function evaluations. Given that the exact shape of the above
integrand can be difficult to predict, keeping $n$ as large as possible was
desirable. Additionally, we again ran into the fact that commodity GPUs achieve
the highest throughput for single precision floating point calculations. This
made the use of mixed-precision necessary, where many of the calculations are
done and variables stored as single precision floats.

Extending Gaussian quadrature in two-dimensions was done simply by setting up a
two-dimensional grid such that
\begin{equation}
\int f(x,y)\mathrm{d}x\mathrm{d}y \cong \sum_{i=1}^{n}\sum_{j=1}^{n}w_{i}w_{j}f(x_{i},x_{j}),
\end{equation}
where $x_{i}$ are the points to evaluate your function determined from your
Gaussian quadrature rule, and $w_{i}$ are their associated weights, both of
which can be readily found in handbooks. Combine the need for a two-dimensional
grid, the unknown shape of the integrand, the need to use mixed precision
and the fact that the maximum number of threads per GPU thread block is 1024,
and $n = 32$ becomes a natural choice.

This allowed us to have one thread block per $k$-triplet, where the integral
was then approximated by a two-dimensional $32\times 32$-point Gaussian
quadrature rule. Each thread then computed one contribution to the integral,
and stored the result in block shared memory, with the final summing done in a
two step reduction. To reduce the impact of mixed precision, we stored all the
calculations of equations \eqref{eq:1stkernel} -- \eqref{eq:Gkernel} as single
precision and the calculations of equation \eqref{eq:bkmod} and \eqref{eq:BispecFoG} as
double-precision. We then perform the final summing over the two-dimensional
grid using those double-precision values and return the result as
double-precision.  In our tests, a complete double-precision calculation using
the exact same algorithm has a relative difference -- e.g. $(B_{\mathrm{MP}} -
B_{\mathrm{DP}})/B_{\mathrm{DP}}$ -- from our mixed-precision calculation of
$\sim$10$^{-7}$. Given the relatively large uncertainties in the measured
bispectrum, this loss of precision was well worth $\sim$12$\times$ speed-up of
the model evaluation.

For the fitting, we used six free parameters: the three from equations
\eqref{eq:1stkernel} and \eqref{eq:2ndkernel}, e.g. $b_1$, linear bias, $b_2$,
second-order bias, and $f$, the linear growth factor\footnote{These parameters
can only be measured up to some overall power spectrum normalization,
$\sigma_{8}$, which we leave off for brevity. In the text, when we use $b_{1}$,
$b_{2}$, or $f$, we mean the combinations $b_{1}\sigma_{8}$, $b_{2}\sigma_{8}$,
or $f\sigma_{8}$} with the two Alcock-Paczynski effect
\citep{Alcock1979,Kaiser1987,Ballinger1996,Simpson2010,Samushia2011}
parameters, $\alpha_{\parallel}$ and $\alpha_{\perp}$, and the Finger-of-God
velocity dispersion parameter, $\sigma_{v}^{2}$. Since our model was only
validated for the purposes of measuring the BAO feature we did not attach any
cosmologically meaningful interpretation to the estimates of the parameters $b_1$,
$b_2$, $f$, or $\sigma_{v}^{2}$. They were reasonably close to the linear model
expectations but were very likely strongly affected by systematics and we
therefore do not quote them as useful cosmological constraints in this work.

\begin{figure}
\includegraphics[width=\linewidth]{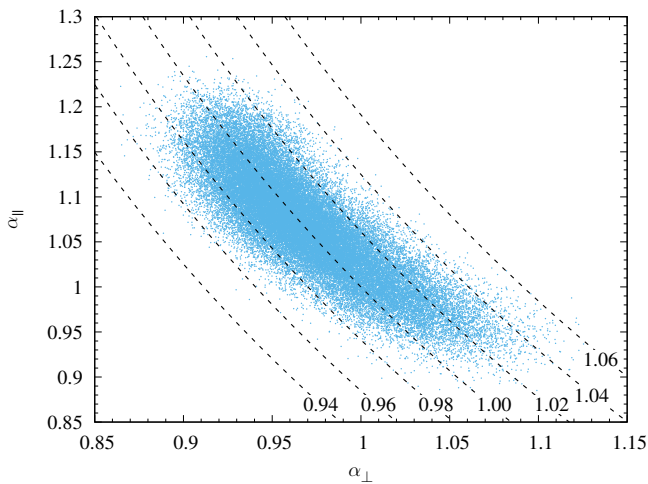}
\caption{Accepted realizations from one of our MCMC chains fitting to the mock
average N+SGC bispectrum projected onto
$\alpha_\parallel$-$\alpha_\perp$ plane. The dashed lines correspond to fixed
values of $\alpha^{3} \equiv \alpha_\perp^2\alpha_\parallel$.}
\label{fig:alphachains}
\end{figure}

Since we were only fitting to the spherically averaged bispectrum monopole, we
didn't expect to be able to reliably constrain both $\alpha_{\parallel}$ and
$\alpha_{\perp}$. Instead, as in the case of the spherically averaged two point
statistics \citep[see e.g.][]{Anderson2012,Anderson2014}, we expected to only
constrain the value of the single scale dilation parameter, $\alpha$.  We
elected to keep the fit in terms of $\alpha_{\parallel}$ and $\alpha_{\perp}$.
Then we calculated the value of $\alpha$ via equation \eqref{eq:alphacombo} for
each of our accepted parameter realizations in the MCMC chain. It is not
immediately obvious that the best constrained combination of $\alpha$s in the
bispectrum is the same as for the power spectrum. By inspecting our MCMC chains
we were able to verify that the two are in fact very close. Figure
\ref{fig:alphachains} shows one of our MCMC chains projected onto the
$\alpha_\parallel$-$\alpha_\perp$ plane and the principle axis of the likelihood
ellipse is very well aligned with the direction of $\alpha$ in equation
\eqref{eq:alphacombo}.

We ran our MCMC chains with no priors aside from the loose requirement the $f$
remain positive. We used our own custom MCMC software utilizing the
Metropolis-Hastings algorithm \citep{Hastings1970}, which was easier to
interface with the model calculation on the GPU. Given that only one GPU was
available for the model calculation and the GPU's constant memory was used to
store the values of the parameters for each model evaluation, we simply ran a
single chain for many millions of realizations instead of running multiple,
simultaneous chains needed to test for convergence. We built our code with the
option to resume a chain should our post-processing reveal that it had not
sufficiently explored the parameter space.

\subsection{The Combined Power Spectrum and Bispectrum Model}

For the joint fit our data vector simply became the 37 power spectrum values,
followed by the 691 bispectrum values. The model was then simply calculating
the first 37 values using the power spectrum model of section \ref{sec:PkMod}
and the next 691 values with the bispectrum model of section
\ref{sec:BispecMod}.

The main difference here came in the free parameters, particularly $\alpha$,
$\alpha_{\perp}$ and $\alpha_{\parallel}$. Instead of letting all three be
free, we instead only let $\alpha_{\perp}$ and $\alpha_{\parallel}$ vary freely
with $\alpha$ for the power spectrum model then being fixed by equation
\eqref{eq:alphacombo}. In all, for the combined model we ended up with 14 free
parameters: $B$, $\Sigma$, $a_{0}$ -- $a_{5}$, $b_{1}$, $b_{2}$, $f$,
$\alpha_{\perp}$, $\alpha_{\parallel}$, and $\sigma_{\mathrm{v}}$. This likely
could have been reduced by relating the power spectrum amplitude parameter $B$
to the biases, $b_{1}$ and $b_{2}$, and the linear growth factor, $f$. However,
since we were only concerned with the constraints on $\alpha$ in the end, all
other parameters were treated as nuisance parameters, and marginalized against
anyway. We again ran our MCMC chains with only loose constraints to ensure that
parameters that enter the models as squares remained positive.

\section{Results}
\label{sec:results}

We performed a number of fittings to the power spectrum, bispectrum and their
combination, analysing the NGC and SGC samples separately before fitting to the
volume weighted averages. This was done to ensure that the results from any one
fitting were consistent with the results from others. We additionally fit the
models to the average power spectrum and bispectrum from the mock catalogues
for the purposes of verifying our analysis pipeline and assessing systematic
errors.

Our fiducial cosmology was virtually identical to that of the MultiDark-\textsc{patchy}
mock catalogues. As such, we would expect that $\alpha$ should have been very
nearly one when fit to the ensemble of mocks -- the exact expected value is
$\alpha_{\mathrm{mock}} = 1.00005$. Looking at Table \ref{tab:alphas}, we can
see that this indeed turned out to be the case. We take the largest per cent
deviation from the expected value of $\alpha$ for each of the fittings -- e.g.
the power spectrum, bispectrum and joint fittings -- to indicate our systematic
error. These errors were $\sim$0.7 per cent for the power spectrum and bispectrum,
and $\sim$0.5 per cent for the combined fitting.

For the DR12 measurements, our results for the NGC and SGC separately are in
agreement with \citet{Ross2017}, who also analysed the NGC and SGC separately,
finding a slightly lower value for the NGC and a slightly larger one for the
SGC, and the combined result being closer to one. We also note the agreement of
our power spectrum and bispectrum results. The largest difference is
for the SGC measurements, which likely had to do with the smaller volume
resulting in a noisier measurement of the bispectrum.

\begin{table}
\caption{The results of the MCMC fittings for the BAO scale parameter,
$\alpha$. Both the values measured from the mocks and the data for the two
galactic cap samples, as well as the volume weight averages, are presented. The
standard deviations have been scaled due to the covariance uncertainty.}
\label{tab:alphas}
\centering
\begin{tabular}{@{}lccc@{}}
\toprule
Data & Sample & $\alpha$ (mock) & $\alpha$ (DR12) \\
\midrule
$P(k)$                        & NGC     & $1.007\pm 0.018$ & $0.976\pm 0.012$ \\
                              & SGC     & $1.003\pm 0.025$ & $1.025\pm 0.021$ \\
                              & N+SGC   & $1.003\pm 0.013$ & $0.988\pm 0.012$ \\ [1ex]
$B(k_{1}, k_{2}, k_{3})$      & NGC     & $1.002\pm 0.020$ & $0.978\pm 0.031$ \\
                              & SGC     & $1.007\pm 0.030$ & $1.086\pm 0.072$ \\
                              & N+SGC   & $1.001\pm 0.017$ & $0.991\pm 0.027$ \\ [1ex]
$P(k)+B(k_{1}, k_{2}, k_{3})$ & NGC     & $1.001\pm 0.016$ & $0.982\pm 0.011$ \\
                              & SGC     & $1.005\pm 0.020$ & $1.020\pm 0.020$ \\
                              & N+SGC   & $1.002\pm 0.010$ & $0.988\pm 0.011$ \\
\bottomrule
\end{tabular}
\end{table}

\begin{figure*}
\includegraphics[width=0.49\linewidth]{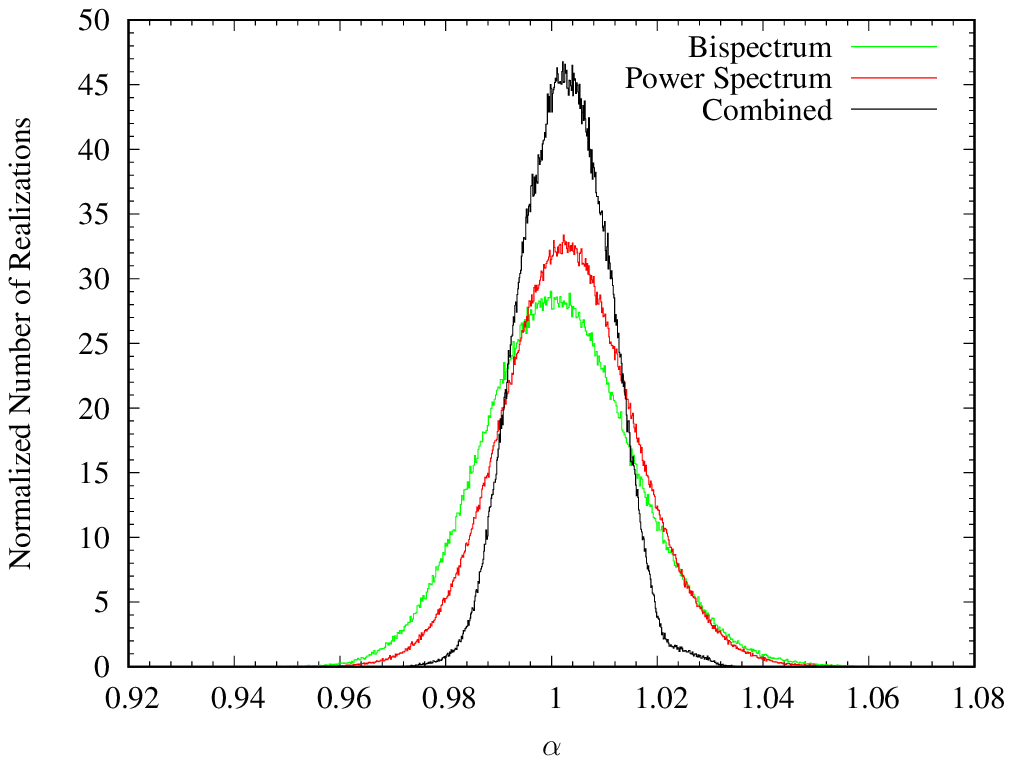}\hfill
\includegraphics[width=0.49\linewidth]{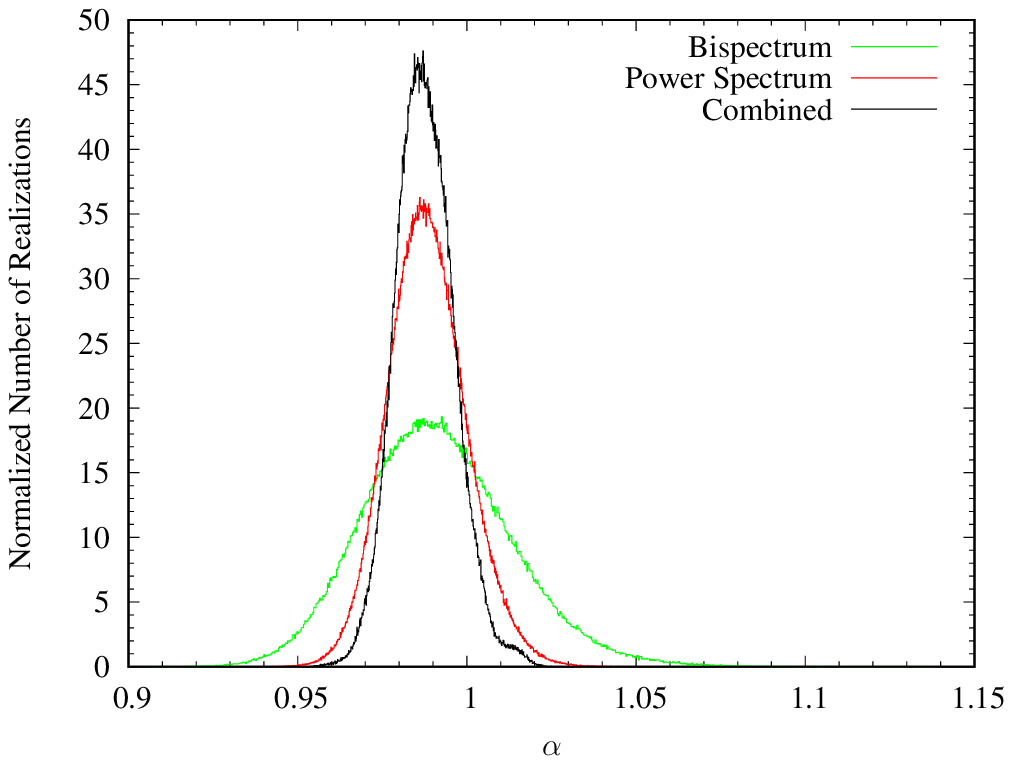}
\caption{Histograms of the MCMC realizations from fitting to the bispectrum
only, the power spectrum only, and the combination for the N+SGC. The left hand
panel shows the results from fitting to the average of the mocks. The right
hand panel shows the results from fitting to the measurments from the data.
(See the online article for a colour version of this plot.)}
\label{fig:alphacomb}
\end{figure*}

\begin{figure*}
\includegraphics[width=0.49\linewidth]{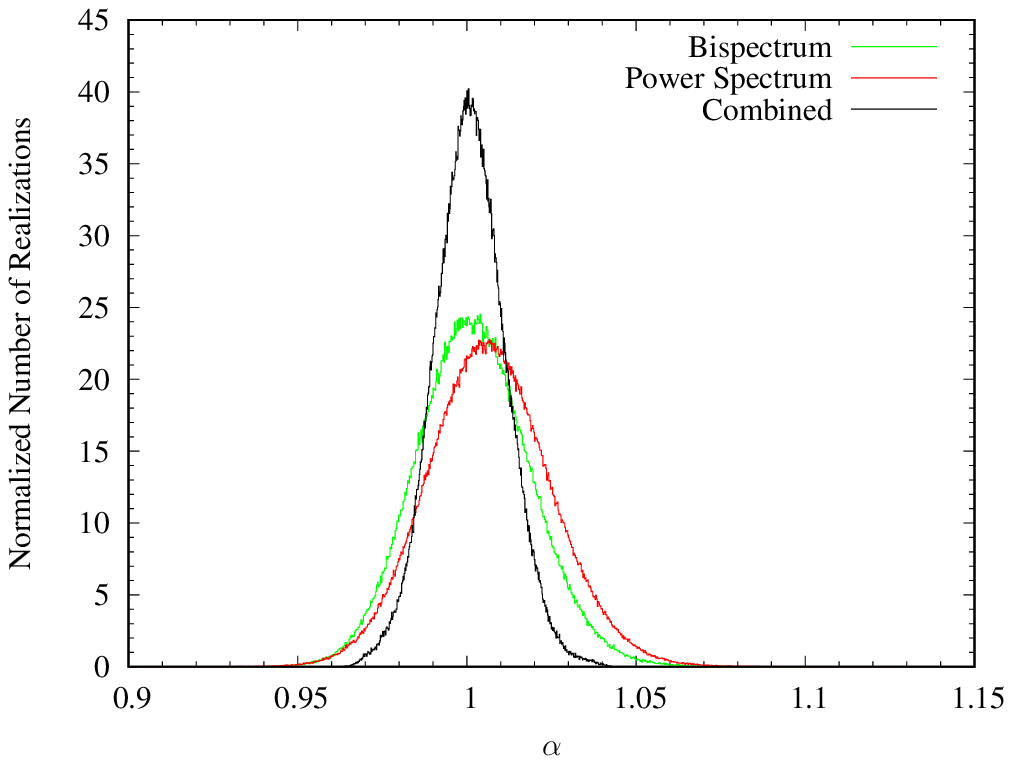}\hfill
\includegraphics[width=0.49\linewidth]{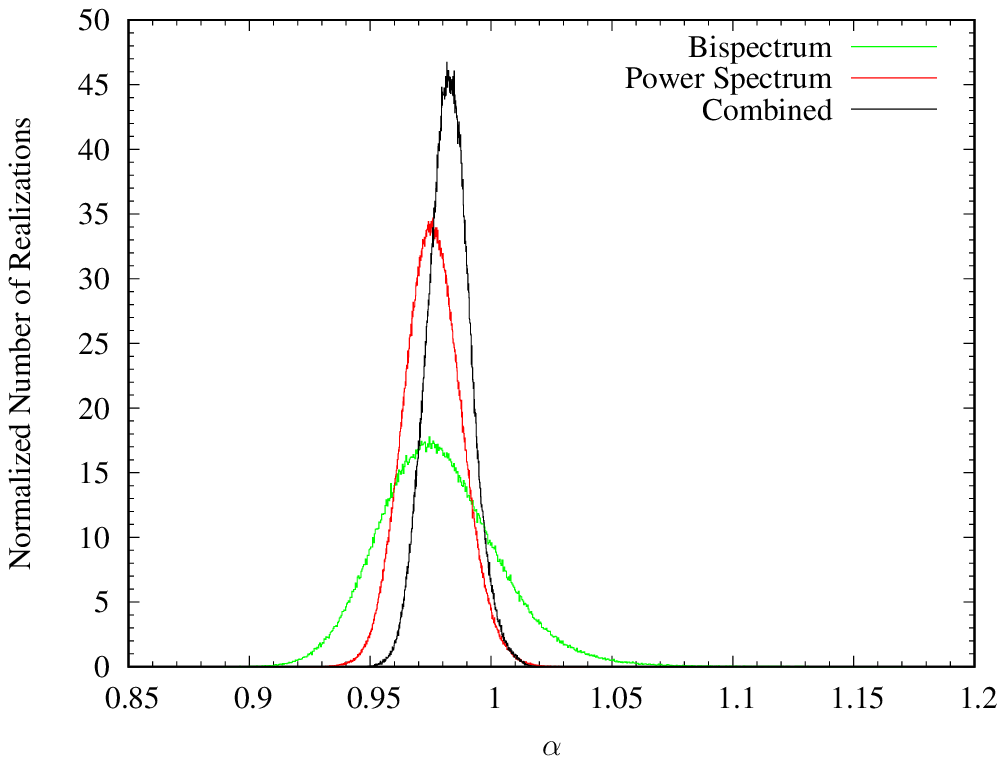}
\caption{Histograms of the MCMC realizations from fitting to the bispectrum
only, the power spectrum only, and the combination for the NGC CMASS sample.
The panels are the same as in Figure \ref{fig:alphacomb}. (See the online
article for a colour version of this plot.)}
\label{fig:alphaNGC}
\end{figure*}

\begin{figure*}
\includegraphics[width=0.49\linewidth]{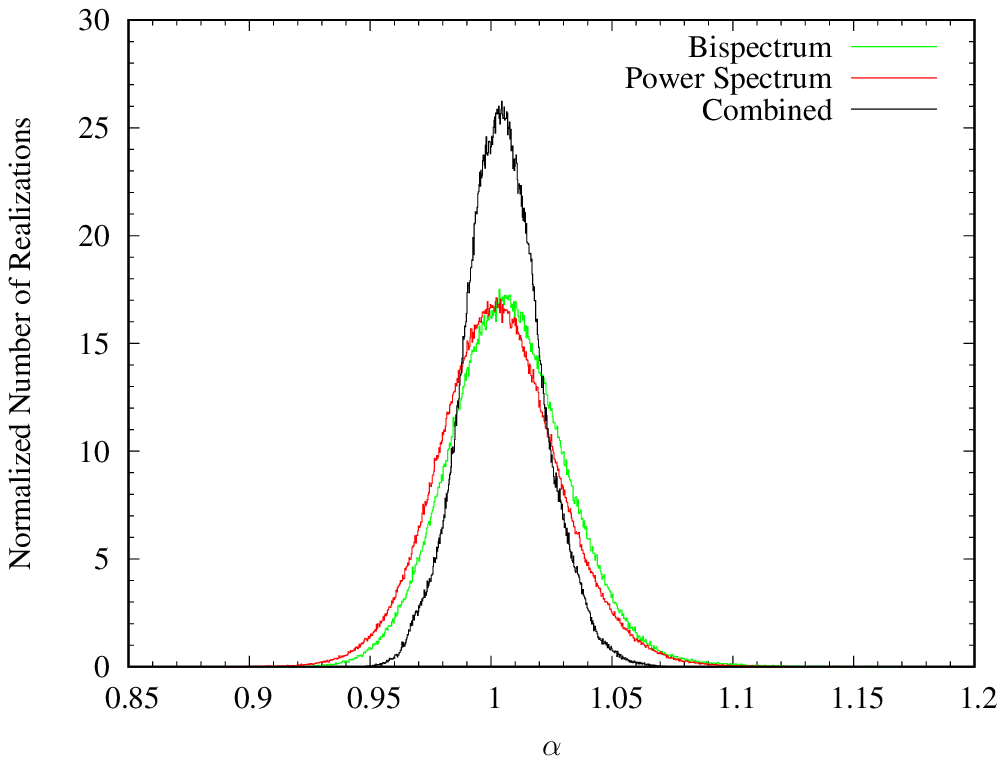}\hfill
\includegraphics[width=0.49\linewidth]{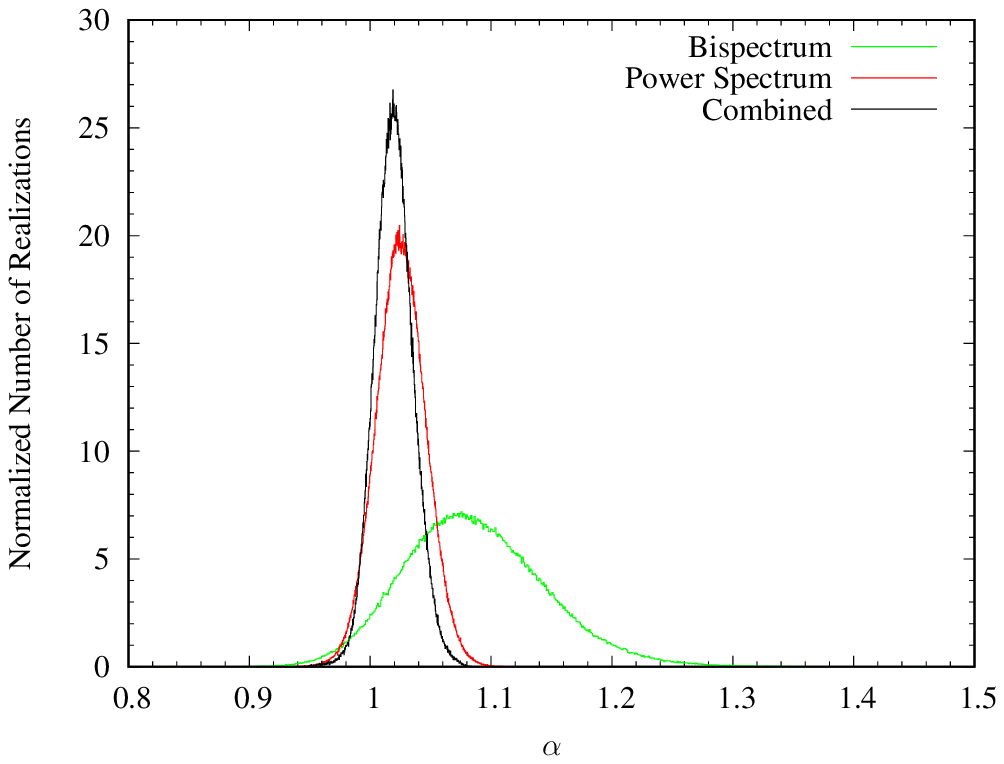}
\caption{Histograms of the MCMC realizations from fitting to the bispectrum
only, the power spectrum only, and the combination for the SGC CMASS sample.
The panels are the same as in Figure \ref{fig:alphacomb}. (See online article
for colour version of this plot.)}
\label{fig:alphaSGC}
\end{figure*}

We show the histograms for $\alpha$ from all of our MCMC fittings in Figures
\ref{fig:alphacomb}, \ref{fig:alphaNGC}, and \ref{fig:alphaSGC} for the N+SGC,
NGC, and SGC samples, respectively. In all of the figures, the results of
fitting to the mocks is shown on the left, and the fitting to the data on the
right, and all histograms have been normalized so that the area under the
curves is equal to one. All of the histograms are very close to Gaussian and
encapsulate all regions of relatively high likelihood suggesting that $\alpha$
parameter space was well explored by our MCMC chains. These plots have not been
broadened by the $m_{1}$ factor of equation \eqref{eq:m1}.

The standard deviations listed in Table \ref{tab:alphas} have been broadened by
the $m_1$ factors. The effect is quite apparent in the full sample, N+SGC,
combined power spectrum-bispectrum fitting. Without carrying the covariance
uncertainty through, the standard deviation would be 0.0087, a $\sim$27 per
cent tighter constraint than the power spectrum alone, and a better constraint
than the one from the power spectrum of the reconstructed field. However, after
multiplying by the square root of $m_{1}$, this become 0.0108 which still
represents a $\sim$10 per cent tighter constraint than our power spectrum fitting.

To test how the constraints may improve given a more precise estimation of the covariance
matrix, we ran
chains fitting to the average of the mocks with the assumption that our covariance 
was drawn from one-million mock catalogues,
which would make the effects both the covariance scaling and $m_{1}$ factor negligible.
We did this for both the bispectrum and joint fitting, finding that the standard
deviation for $\alpha$ dropped to 0.011 and 0.007 from 0.014 and 0.010, respectively.
This indicates that given a less noisy estimate of the covariance, from either more
mock catalogues, a high-precision theoretical calculation, or a hybrid approach like
the one used by \citet{Slepian2017a,Slepian2017b}, could improve constraints
on the distance scale by 30 per cent.

We note that we achieve a significantly better fit by using a `wiggle'
power spectrum in our bispectrum model than a `no-wiggle' power spectrum, with a $\chi^2$ penalty
for the no-wiggle model of $\Delta\chi^{2} = 20.64$. This implies a
$4.1\sigma$ detection of the BAO features in the BOSS DR12 galaxy bispectrum,
comparable to the significance of the detection by \citet{Slepian2017b}.

Of course, $\alpha$ itself is merely a means to measure the distance to the
survey redshift via equation \eqref{eq:DVmeas}. First it was necessary to
compute $D_{\mathrm{V}}^{\mathrm{fid}}(z)$, which requires calculating
$D_{\mathrm{A}}^{\mathrm{fid}}(z)$, and $H^{\mathrm{fid}}(z)$. For our fiducial
cosmology, these come out to $H^{\mathrm{fid}}(z = 0.57) =
93.04~\mathrm{km}~\mathrm{s}^{-1}~\mathrm{Mpc}^{-1}$,
$D_{\mathrm{A}}^{\mathrm{fid}}(z = 0.57) = 1386.01~\mathrm{Mpc}$, and
$D_{\mathrm{V}}^{\mathrm{fid}}(z = 0.57) = 2056.45~\mathrm{Mpc}$. We also note
that the drag radius for our fiducial cosmology was
$r_{\mathrm{d}}^{\mathrm{fid}} = 147.59~\mathrm{Mpc}$ as calculated from the
fitting formula of \citet{Hu1996} and \citet{Eisenstein1998}.  The values of
$D_{\mathrm{V}}$ for the various fittings are reported in Table \ref{tab:DVs},
where the $r_{\mathrm{d}}^{\mathrm{fid}}/ r_{\mathrm{d}}$ was omitted from the
column headings for brevity.

\begin{table}
\caption{The $D_{\mathrm{V}}$ values from the various fittings in Table
\ref{tab:alphas}. While the columns are labeled as $D_{\mathrm{V}}$ we note
that they actually represent $(r_{\mathrm{d}}^{\mathrm{fid}}/
r_{\mathrm{d}})D_{\mathrm{V}}$ values. The values for DR12 reported here agree
remarkably well with the values reported by other analyses, and the values from
the fittings to the mocks agree with our expectations.}
\label{tab:DVs}
\centering
\begin{tabular}{@{}lccc@{}}
\toprule
Data & Sample & $D_{\mathrm{V}}$ (mock) & $D_{\mathrm{V}}$ (DR12) \\
     &        & (Mpc)                   & (Mpc)                   \\  
\midrule
$P(k)$                        & NGC     & $2070\pm 38$ & $2007\pm 24$ \\
                              & SGC     & $2063\pm 51$ & $2108\pm 42$ \\
                              & N+SGC   & $2063\pm 26$ & $2032\pm 24$ \\ [1ex]
$B(k_{1}, k_{2}, k_{3})$      & NGC     & $2060\pm 42$ & $2012\pm 63$ \\
                              & SGC     & $2071\pm 62$ & $2233\pm 148$ \\
                              & N+SGC   & $2059\pm 36$ & $2038\pm 55$ \\ [1ex]
$P(k)+B(k_{1}, k_{2}, k_{3})$ & NGC     & $2058\pm 27$ & $2020\pm 23$ \\
                              & SGC     & $2066\pm 41$ & $2097\pm 41$ \\
                              & N+SGC   & $2061\pm 21$ & $2031\pm 22$ \\
\bottomrule
\end{tabular}
\end{table}

We take the measurement from the combined power spectrum plus bispectrum
fitting to the N+SGC data as our main result,
\begin{equation}
D_{\mathrm{V}}(z = 0.57) = 2031 \pm 22~\mathrm{Mpc}~\left(\dfrac{r_{\mathrm{d}}}
{r_{\mathrm{d}}^{\mathrm{fid}}}\right).
\end{equation}
We compared our result to that of other works to test its robustness, after
converting those results to be consistent with our fiducial cosmology. 
\citet{Anderson2014} found
\begin{equation}
D_{\mathrm{V}}^{\mathrm{A14}}(z = 0.57) = 2032\pm 20~\mathrm{Mpc}~\left(\dfrac{r_{\mathrm{d}}}
{r_{\mathrm{d}}^{\mathrm{fid}}}\right),
\end{equation}
from their analysis of the DR11 CMASS sample. In analysing the DR12 CMASS
sample \citet{Cuesta2016} found
\begin{equation}
D_{\mathrm{V}}^{\mathrm{C16}}(z = 0.57) = 2035\pm 20~\mathrm{Mpc}~\left(\dfrac{r_{\mathrm{d}}}
{r_{\mathrm{d}}^{\mathrm{fid}}}\right),
\end{equation}
\citet{Gil-Marin2016} found
\begin{equation}
D_{\mathrm{V}}^{\mathrm{G16}}(z = 0.57) = 2035\pm 19~\mathrm{Mpc}~\left(\dfrac{r_{\mathrm{d}}}
{r_{\mathrm{d}}^{\mathrm{fid}}}\right),
\end{equation}
\citet{Ross2017} found,
\begin{equation}
D_{\mathrm{V}}^{\mathrm{R17}}(z = 0.57) = 2022\pm 17~\mathrm{Mpc}~\left(\dfrac{r_{\mathrm{d}}}
{r_{\mathrm{d}}^{\mathrm{fid}}}\right),
\end{equation}
and \citet{Slepian2017b} found
\begin{equation}
D_{\mathrm{V}}^{\mathrm{S17m}}(z = 0.57) = 2036\pm 33~\mathrm{Mpc}~\left(\dfrac{r_{\mathrm{d}}}
{r_{\mathrm{d}}^{\mathrm{fid}}}\right),
\end{equation}
from their `minimal' model and 
\begin{equation}
D_{\mathrm{V}}^{\mathrm{S17t}}(z = 0.57) = 2026\pm 29~\mathrm{Mpc}~\left(\dfrac{r_{\mathrm{d}}}
{r_{\mathrm{d}}^{\mathrm{fid}}}\right),
\end{equation}
from their `tidal' model. Our main result deviates the most from that of
\citet{Ross2017}, and even then the disagreement is within $\sim$0.5$\sigma$.

Additionally, \citet{Ross2017} gave results for the NGC and SGC separately,
allowing for a more detailed comparison. Their fitting to the two-point
correlation function gave
\begin{equation}
D_{\mathrm{V}}^{\mathrm{R17,NGC}}(z = 0.57) = 2006\pm 21~\mathrm{Mpc}~\left(\dfrac{r_{\mathrm{d}}}
{r_{\mathrm{d}}^{\mathrm{fid}}}\right),
\end{equation}
and 
\begin{equation}
D_{\mathrm{V}}^{\mathrm{R17,SGC}}(z = 0.57) = 2090\pm 41~\mathrm{Mpc}~\left(\dfrac{r_{\mathrm{d}}}
{r_{\mathrm{d}}^{\mathrm{fid}}}\right).
\end{equation}
These values agree remarkably well with our analysis, showing a somewhat lower
value from the NGC and a higher value from the SGC.

Similar to \citet{Slepian2017b}, we assessed our systematic errors by examining
the bias of our results of fitting to the mocks. We take the largest deviation
from the expected value of $\alpha$ in each of the fittings, finding systematic
errors of $\pm15$~Mpc for the power spectrum and bispectrum fittings, and 
$\pm10$~Mpc for the combined fitting. This gives our main result as
\begin{equation}
D_{\mathrm{V}}(z = 0.57) = 2031 \pm 22~\mathrm{(stat.)}\pm 10~\mathrm{(sys.)}~\mathrm{Mpc}~
\left(\dfrac{r_{\mathrm{d}}}
{r_{\mathrm{d}}^{\mathrm{fid}}}\right).
\end{equation}

\section{Conclusions}
\label{sec:conclusion}

We report an independent measurement of the distance to the BOSS DR12 CMASS
sample of 
$D_{\mathrm{V}}(z=0.57) = 2032 \pm 24~\mathrm{Mpc}~(r_{\mathrm{d}}/r_{\mathrm{d}}^{\mathrm{fid}})$ from
the power spectrum, 
$D_{\mathrm{V}}(z=0.57) = 2038 \pm 55~\mathrm{Mpc}~(r_{\mathrm{d}}/r_{\mathrm{d}}^{\mathrm{fid}})$ from 
the bispectrum, and 
$D_{\mathrm{V}}(z=0.57) = 2031 \pm 22~\mathrm{Mpc}~(r_{\mathrm{d}}/r_{\mathrm{d}}^{\mathrm{fid}})$ from
the combined analysis. These values are in remarkable agreement with each other, and with the analyses of
\citet{Anderson2014}, \citet{Cuesta2016}, \citet{Gil-Marin2016}, \citet{Ross2017} and \citet{Slepian2017b}.
The power spectrum gives a $\sim$1.2 per cent constraint ($\sim$1.4 per cent with systematics), 
and the bispectrum
gives a $\sim$2.7 per cent constraint ($\sim$2.8 per cent with systematics). The combined analyses
gives a $\sim$1.1 per cent constraint ($\sim$1.2 per cent with systematics), mainly limited by
the number mocks available for covariance estimation.
However, when combined the constraint still improves by
$~\sim$10 per cent compared to the power spectrum only constraints.

Our bispectrum constraints from the mocks were tighter than the ones from the
data, suggesting that this specific realisation of the DR12 CMASS volume is
slightly noisier than typical from the point of view of the bispectrum monopole
estimator. When fitting to the mean of the mocks we get a 1.7 per cent
constraint from the bispectrum only and a 1 per cent constraint from the joint
fit to the power spectrum and bispectrum. The numbers in table \ref{tab:alphas}
suggest that if a better model for the covariance were available the joint
constraints from the power spectrum and the bispectrum would be comparable to
the constraints from reconstructed power spectrum even at the current level of
systematics.

The main limiting factor to the precision of the bispectrum measurements is a
relatively small number of mock catalogues available for the covariance
estimation. Having only 2048 mocks means that the values in our inverse
covariance matrix estimate were all substantially reduced -- by a factor of
0.662 for the bispectrum and 0.644 for the combined data -- leading to broader
posterior likelihoods. Having a better estimate of the bispectrum and joint
covariance matrices would reduce the error on the $\alpha$ parameter from the
joint fit by an extra $\sim$30 per cent. These improved estimates could come from
either a larger number of mocks or analytic calculations.

\begin{figure}
\includegraphics[width=\linewidth]{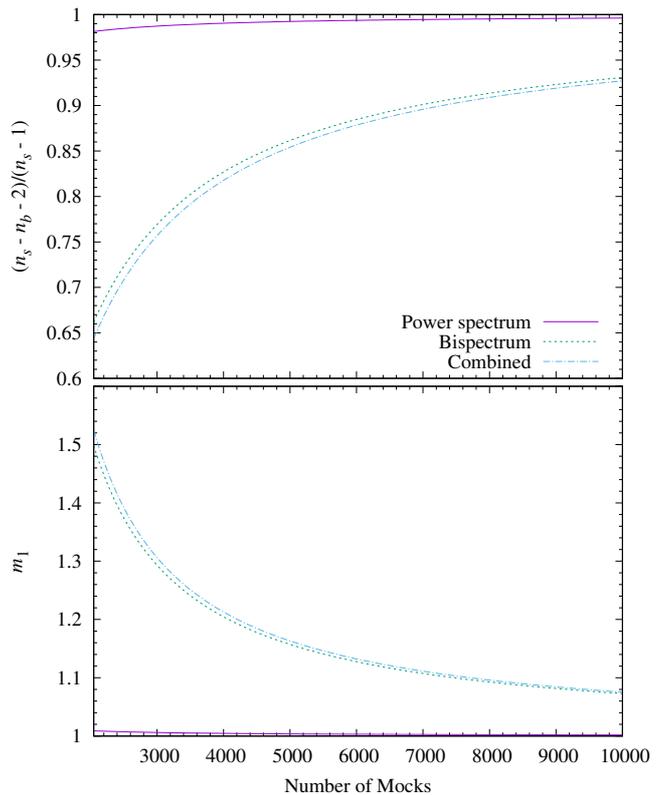}
\caption{The effects of the limited number of mock catalogues. The top panel
shows the factor that multiplies the inverse sample covariance, and the bottom
panel shows how much the variance of parameter constraints should be inflated.
The solid line (purple online) shows the quantities for the power spectrum, the
dashed line (green online) shows them for the bispectrum, and the dot-dashed
line (cyan online) shows them for the combined data vector. The
horizontal axis starts at the number of mocks used in this work. It is clear
that increasing the number of mocks will have little effect on the power
spectrum, but a dramatic effect on bispectrum and combined data.}
\label{fig:numMocks}
\end{figure}

Looking at Figure \ref{fig:numMocks}, we can clearly see that the power
spectrum constraints will not really benefit from an increased number of mock
catalogues, while the constraints from the bispectrum and the combined data can
see dramatic improvements with $\sim$10~000 mocks. With $\sim$40~000 mocks,
the additional uncertainty due to noise in the inverse covariance matrix would
become negligible.

However, generating that number of mock catalogues is a computationally
expensive proposition, which is only going to be exacerbated by the increased
volumes and number densities of future redshift surveys such as, the upcoming
DESI \citep{Schlegel2011,Levi2013,DESI}
survey, Large Synoptic Survey Telescope \citep[LSST;][]{LSST2009} surveys,
\textit{Euclid} satellite mission surveys \citep{Laureijs2011}, and \textit{WFIRST}
\citep{Green2012} surveys.
\citet{Monaco2016} estimates that for these future surveys, generating 1000
mock realizations with the fastest of the popular mock codes currently in use
would take $\sim$1~000~000 CPU hours since, unfortunately, the cheapest mock
catalogues to produce, the lognormal mocks \citep*{Coles1991,Beutler2011,
Pearson2016b}, do not adequately reproduce the
three-point statistics \citep[see][Figure 9]{White2014}.

It would also be interesting to test if various methods of reducing the
uncertainties in the covariance matrix such as shrinkage estimation
\citep{Pope2008}, estimation from fitting formula \citep{Pearson2016}, or
calculating the expected covariance from theory \citep{Xu2012}, could work for
the bispectrum. \citet{Slepian2017a,Slepian2017b} used a hybrid approach, fitting a
theoretical model to the sample covariance from a 299 mocks, which may also be
a useful approach for the bispectrum.

Lastly, although we find our theoretical template to be unbiased for this
analysis, it would be interesting to test if using a more complex bispectrum
model, such as the one used by \citet{Gil-Marin2015a} and
\citet{Gil-Marin2017}, would affect the constraints. We leave these matters for
exploration in future works.

\section*{Acknowledgements}
LS is grateful for the support from SNSF grant SCOPES IZ73Z0 152581, GNSF grant
FR/339/6-350/14, NASA grant 12-EUCLID11-0004, and the DOE grant DE-SC0011840.
NASA's Astrophysics Data System Bibliographic Service and the arXiv e-print
service were used for this work. Additionally, we wish to acknowledge
\textsc{gnuplot}, a free open-source plotting utility which was used to create
all of our figures.




\bibliographystyle{mnras}
\bibliography{bispec} 

\bsp	
\label{lastpage}
\end{document}